# Anisotropic resistivity and Hall effect in MgB$_2$ single crystals

Yu. Eltsev, K. Nakao, S. Lee, T. Masui, N. Chikumoto, S. Tajima, N. Koshizuka, and M. Murakami

Superconductivity Research Laboratory, ISTEC,
10-13, Shinonome 1-chome, Koto-ku, Tokyo, 135-0062, Japan

*We report resistivity and the Hall effect measurements in the normal and superconducting states of MgB$_2$ single crystal. The resistivity has been found to be anisotropic with slightly temperature dependent resistivity ratio of about 3.5. The Hall constant, with a magnetic field parallel to the Mg and B sheets is negative in contrast to the hole-like Hall response with a field directed along the c-axis indicating presence of both types of charge carriers and, thus, multi-band electronic structure of MgB$_2$. The Hall effect in the mixed state shows no sign change anomaly reproducing the Hall effect behavior in clean limit type-II superconductors.*

Following the recent discovery of superconductivity at about 39K in magnesium diboride [1] various properties of this compound have been extensively studied. Early observation of boron isotope effect [2] clearly indicated an important role of the electron-phonon interaction in $MgB_2$. Subsequent measurements of specific heat [3], $^{11}$B NMR [4], and Raman scattering [5] provided evidence for s-wave order parameter symmetry. However, more recent studies of quasiparticle tunneling [6,7], specific heat [8-10], and penetration depth [11] revealed some unusual features indicating double-energy superconducting gap structure. Theoretical studies also give support for this scenario. In particular, the existence of two distinct superconducting gaps in $MgB_2$ with a smaller one on the 3D-tubular network and a larger one on the 2D sheets was predicted by Liu *et al.* from the first-principles calculations [12].

In spite of quite complex electronic band structure of $MgB_2$, results of theoretical calculations [12-14] and direct studies of the Fermi surface in angle resolved photoemission [15] and de Haas-van Alphen [16] experiments are in a good agreement. For further understanding of the electronic structure of $MgB_2$ it is crucially important to know the nature of charge carriers in this compound. Hall effect measurement is one of the powerful tools to obtain such information. Recently the Hall effect in $MgB_2$ has been studied using hot-pressed polycrystalline [17] and thin film [18,19] samples. All these experiments gave a positive (hole-like) sign of the Hall effect and about one order of magnitude lower value of the normal state Hall constant, compared to the case for conventional superconductors, like $Nb_3Sn$ and $Nb_3Ge$. However, the other features of the reported Hall effect of $MgB_2$ are rather contradictory. In particular, temperature dependence of the Hall constant was different among the reported data [17-19]. Also, in the measurements below $T_c$ Jin *et al.* [18] observed a sign change of the Hall resistivity before it reaches zero, while Kang *et al.* found no sign-reversal in their study of the Hall effect in the mixed state [19].

The anisotropy of the upper critical field of $MgB_2$ is well established now [20-22]. Availability of $MgB_2$ single crystals [21] opens unique opportunity to study the anisotropy of the electrical transport properties of this compound. Here we report measurements of the in-plane, $\rho_{xx} \equiv \rho_{ab}$, and the out-of-plane, $\rho_{zz} \equiv \rho_c$, resistivity of $MgB_2$ single crystal as well as the Hall effect with magnetic field applied parallel to the c-axis and along the ab-planes in both the normal and superconducting states.

Magnesium diboride single crystals have been grown in quasi-ternary $Mg-MgB_2-BN$ system under 4-6 GPa pressure at 1400-1700$^o$C, as described previously [21]. Several plate-like single crystals of dimensions of ~0.5x0.1x0.03mm³ have been selected for the in-plane transport measurements while in experiments with current along the c-axis we used thicker crystals of ~0.2x0.1x0.1mm³ in size. All the crystals have $T_c$, defined as the resistivity onset with criterion of 2% of the full resistivity drop, of about 38.8K with a transition width within 0.2-0.3K. Stable, low-resistance (~1-2 ) electrical contacts were made using gold or silver paste without subsequent heat treatment. For the measurements of the anisotropic resistivity, the contact configuration with two contacts on both ab-planes of a crystal (sample 1) was used as shown in the inset of Fig. 1, and $\rho_{ab}$ and $\rho_c$ values were obtained from the Montgomery-type analysis. In our Hall effect experiments, the Hall resistivity, $\rho_{ij}$, was extracted from the antisymmetric part of the transverse voltage response under magnetic field reversal, and the Hall constant, $R^H_{ij}$, was calculated as $R^H_{ij}= \rho_{ij}/H$. The in-plane Hall effect measurements (sample 2) have been performed with homogeneous current parallel to the ab-planes and magnetic fields directed along the c-axis. In the out-of-plane Hall effect study both the out-of-plane Hall resistivity components, $\rho_{xz}$ and $\rho_{zx}$, were obtained from the measurements in magnetic field //ab-planes and homogeneous current //ab-planes (sample 3) and //c-axis (sample 4), respectively. The experiments have been carried out in a Quantum Design Physical Property Measurement System in magnetic fields up to 9T with a sample mounted on a horizontal rotator. Current-voltage response was recorded using usual low frequency (17Hz) ac-technique with excitation current of 0.5-1mA within a linear regime.

Anisotropic behavior of resistivity of $MgB_2$ single crystal is illustrated in Fig. 1. The inset of Fig. 1 presents raw data obtained from the two sets of measurements with current injected along the ab-planes ($I_{12}$) and //c-axis ($I_{13}$). Due to the combined effect of the contact configuration, crystal geometry and resistivity anisotropy values of $R_1=V_{34}/I_{12}$ and $R_2=V_{24}/I_{13}$ are rather different with slightly temperature dependent ratio $R_1/R_2$ 6. To extract values of the in-plane and the out-of-plane resistivity, we used the Montgomery-type analysis [23,24]. Since the electrical terminals cover the significant portion of a sample surface, the finite contact size was taken into account. In our calculations we assumed the uniform current injection within the contacts area. We also supposed that the voltage contacts do not disturb the current distribution within the sample and, thus, probe the averaged value of the potential at the interface between the sample and voltage terminal. Both assumptions are safely justified by the high conductivity of the crystal compared to the contact resistance.

Obtained from our analysis results for $\rho_{ab}(T)$ and $\rho_c(T)$ are shown in the main panel of

Fig. 1. Very close agreement between resistivity data for sample 1 and sample 2 (calculated from the Montgomery-type analysis and directly measured with homogeneous in-plane current distribution, respectively) gives a proof of a validity of our analysis [25]. From Fig. 1 one can see pronounced resistivity anisotropy of $MgB_2$ single crystal. Just above $T_c$ we obtain resistivity ratio $\rho_c/\rho_{ab}=3.6\pm1.0$. Rather small and monotonous decrease of this ratio with temperature, down to about 3.4 at T=273K, suggests similar temperature dependence of $\rho_{ab}$ and $\rho_c$, and, thus, the same scattering mechanism for the in-plane as well as the out-of-plane charge transport. Actually, both $\rho_{ab}(T)$ and $\rho_c(T)$ for $MgB_2$ may be fairly well described by a Bloch-Grüneisen expression for resistivity [18]

$$\rho = \rho_0 + C (4\pi)^2 (2T/\Theta_D)^5 \int_0^{\Theta_D/2T} x^5/\sinh^2(x)dx, \quad (1)$$

where $\Theta_D$ is the Debye temperature, $\rho_0$ is the residual resistivity, and C is a proportionality constant. The dotted lines in Fig. 1 represent Bloch-Grüneisen behavior as obtained from Eq. (1) with the same $\Theta_D=880K$ [26], and different values of the residual resistivity of $0.69\mu\Omega$ cm and $2.62\mu\Omega$ cm, and constant C of $0.25\mu\Omega$ cm and $0.86\mu\Omega$ cm for $\rho_{ab}$ and $\rho_c$, respectively. Thus, in an agreement with theoretical calculations of the electronic structure and electron-phonon interaction in $MgB_2$ [12-14] present result demonstrates importance of the electron-phonon scattering for both the in-plane and the out-of-plane resistivity.

A striking anisotropy is observed in the normal state Hall effect. Shown in Fig. 2 are the in-plane and the out-of-plane Hall constant as a function of temperature, obtained from the linear dependence of the Hall resistivity on magnetic field up to 9T at various temperatures. As in the previous studies [17-19], the in-plane Hall constant has been found to be positive, while the out-of-plane Hall response with H//ab-planes shows n-type of charge carriers. Our finding of both types of charge carriers clearly indicates $MgB_2$ as a multi-band metal, in an agreement with calculations of the band structure predicting that the Fermi surface of this compound consists of two quasi two-dimensional hole sheets and two three-dimensional light hole and electron honeycombs [12-14], In fact, in two parabolic band model with both electrons and holes as charge carriers, the Hall constant is a sum of the contributions from each band

$$R^H = (p\mu_p^2 - n\mu_n^2)/ec(p\mu_p + n\mu_n)^2, \quad (2)$$

where e is the carrier charge, c is light velocity, n and p are the densities of electrons and holes, respectively, and $\mu_n$ and $\mu_p$ are the corresponding mobilities [27]. From Eq. (2), positive sign of the in-plane Hall constant found in our measurements may be obtained supposing larger hole-term, while the opposite sign of the out-of-plane Hall constant corresponds to larger electronic contribution. Temperature dependent behavior of both the in-plane and the out-of-plane Hall constants found in our experiment (see Fig. 2) is also expected within a two-band model if one takes into account a possibility of different temperature dependence of the hole and electron terms in Eq. (2). Assuming no contribution from the electrons and holes to the in-plane and the out-of-plane electrical transport, respectively, from Eq. (2) we can estimate the upper limit of electron and hole density as $n\sim 3.4\times 10^{22} cm^{-3}$ and $p\sim 2.6\times 10^{22} cm^{-3}$ at T=40K. These values are about one order of magnitude lower than the previously reported carrier concentration for polycrystalline $MgB_2$ samples [17-19] and rather close to the carrier density in $Nb_3Ge$ and $Nb_3Sn$ [28].

We also would like to point out, that different signs of the in-plane and the out-of-plane Hall constant in $MgB_2$ resemble the anisotropic Hall effect in $YBa_2Cu_3O_{7-}$ with p- and n-type of the Hall response obtained in studies with magnetic field applied //c-axis and //ab-planes, respectively [29,30]. However, this similarity does not extend to the temperature dependence of the Hall effect in $MgB_2$ and high-$T_c$ superconductors. In particular, measurements of the Hall effect in various high-$T_c$ compounds revealed anomalously strong temperature dependence of the in-plane Hall constant that can be turned into simple $T^2$ dependence of the cotangent of the Hall angle, $\cot_H = \rho_{xx}/\rho_{xy}$ (for a review, see [31]). According to Anderson [32], this unusual behavior of the Hall effect in high-$T_c$ cuprates was interpreted as a result of the existence of two distinct relaxation time scales. On the contrary, for $MgB_2$, $\cot_H(T)$ does not demonstrate $T^2$ temperature dependence as obtained from our resistivity and the Hall effect measurements (not shown). Furthermore, the present observation of two types carrier behavior seems to be well supported by the band theory considering $MgB_2$ as a conventional multi-band metal.

Finally, we present results of the Hall effect measurements in the mixed state of $MgB_2$ single crystal. Shown in Fig. 3 are temperature dependences of the in-plane longitudinal and Hall resistivity (top and bottom panel, respectively) measured in magnetic fields //c-axis up to 5T. As reported previously [22], $\rho_{xx}$ exhibits remarkable broadening of the superconducting transition in a magnetic field. Simultaneously measured $\rho_{xy}$ has been found to be positive in the entire transition region and with decreasing temperature monotonously decrease to zero. Data for the out-of-plane Hall effect (not shown) display a very similar behavior, except for negative sign of the out-of-plane Hall constant, smaller transition width and higher transition temperature in a magnetic field of the same magnitude compared to the results of Fig. 3. Although the anomalous Hall constant sign reversal was previously reported for $MgB_2$ polycrystalline films [18], we find no sign change of the Hall effect in both magnetic field orientations. This striking difference does not look very much surprising, since in polycrystalline samples the Hall coefficient could represent a sum of the in-plane and the out-of-plane Hall responses of opposite polarities.

In conventional superconductors, the Hall effect sign change anomaly near $T_c$ was reported for moderately clean superconductors with ratio $l/\xi_0=0.5\div5$, where $l$ is the electron mean-free path, and $\xi_0$ is a coherence length, while sign reversal does not occur in either clean ($l\gg\xi_0$) or dirty ($l\ll\xi_0$) limits (see [33], and references therein). For $MgB_2$, the in-plane and the out-of-plane coherence length were found of about 68Å and 23Å, respectively [22]. Using data for anisotropic resistivity and carrier concentration obtained from our normal state measurements, and the Fermi velocity of $4.9\times10^7$cm/s //ab-planes and of $4.76\times10^7$cm/s along the c-axis [13], we estimate corresponding electron mean-free path as ~700Å and ~180Å just above $T_c$, and, thus, $l/\xi_0\sim10$ for both directions. From this consideration, we can conclude that the absence of the Hall effect sign change in $MgB_2$ is in a nice agreement with the empirical correlation between microscopic material parameters and the mixed state Hall effect behavior reported for type-II superconductors [33].

In summary, the in-plane and the out-of-plane electrical transport properties of $MgB_2$ single crystal have been studied. We found substantial resistivity anisotropy with resistivity ratio $\rho_c/\rho_{ab}$ of about 3.5. Both $\rho_{ab}$ and $\rho_c$ display similar temperature dependence described by a Bloch-Grüneisen expression as a first approximation, thus, indicating predominant contribution of the electron-phonon interaction to the in-plane as well as the out-of-plane charge transport in $MgB_2$. Measurements of the normal state Hall effect with magnetic field applied parallel to the ab-planes and to the c-axis revealed presence of both types of charge carriers. In agreement with the theoretical prediction, this result clearly demonstrates the multi-band electronic structure of $MgB_2$. In the mixed state, the in-plane as well as the out-of-plane Hall constant display no sign change anomaly reproducing the Hall effect behavior in clean limit type-II superconductors.

This work was supported by the New Energy and Industrial Technology Development Organization (NEDO) as Collaborative Research and Development of Fundamental Technologies for Superconductivity Applications.

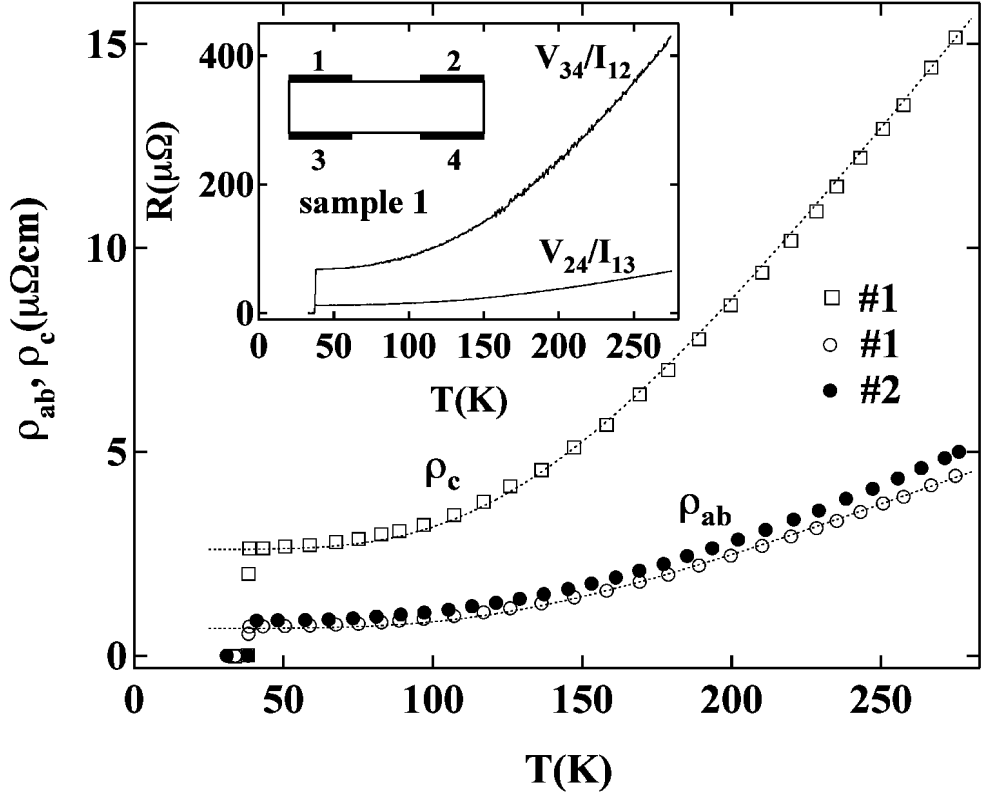

**Fig. 1.** Zero-field temperature dependence of the in-plane and the out-of-plane resistivity of MgB$_2$ single crystal deduced from the Montgomery type analysis for sample 1. For comparison, $_{ab}$(T) dependence obtained from the measurements with homogeneous in-plane current on sample 2 is also shown. Dotted lines represent Bloch-Grüneisen behavior as obtained from Eq.(1). Inset: Definition of contact arrangement and raw current-voltage data for sample 1.

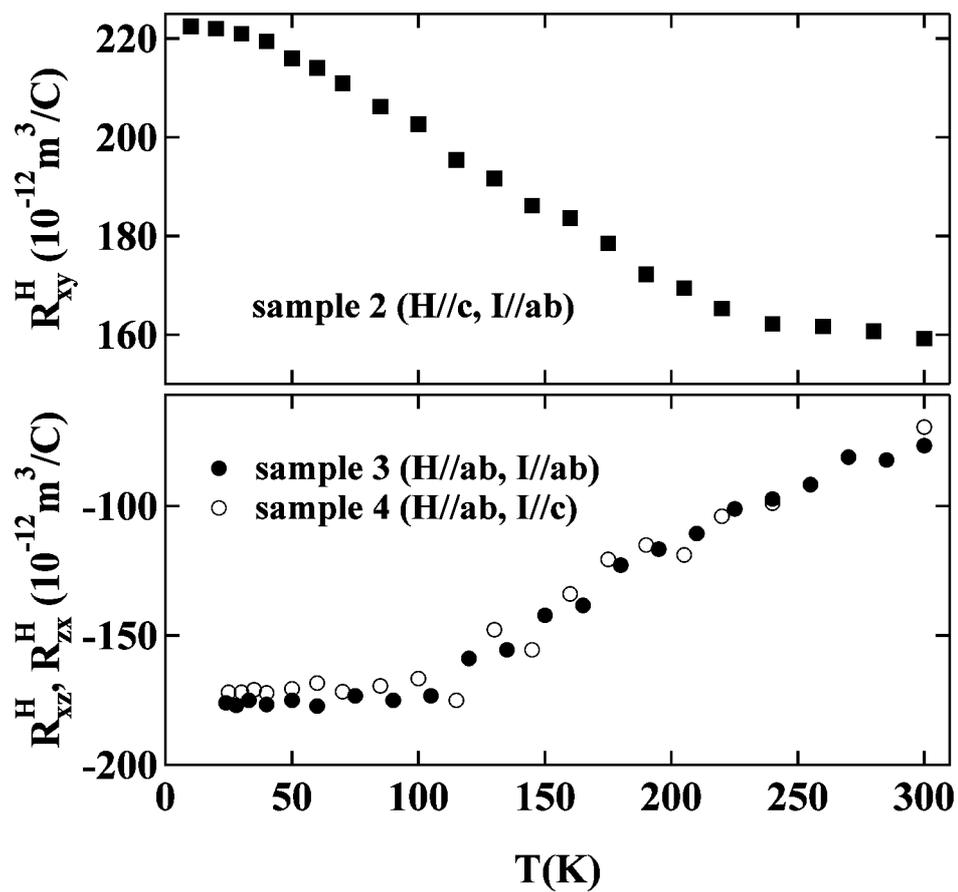

**Fig. 2.** The in-plane and the out-of-plane Hall constant, as a function of temperature in the normal state of $MgB_2$ single crystals (top and bottom panels respectively). The out-of-plane Hall response was measured on two crystals with current parallel to the ab-planes (sample 3) and I//c-axis (sample 4). In accordance with the Onsager relation, the data for both samples demonstrate close agreement within experimental error.

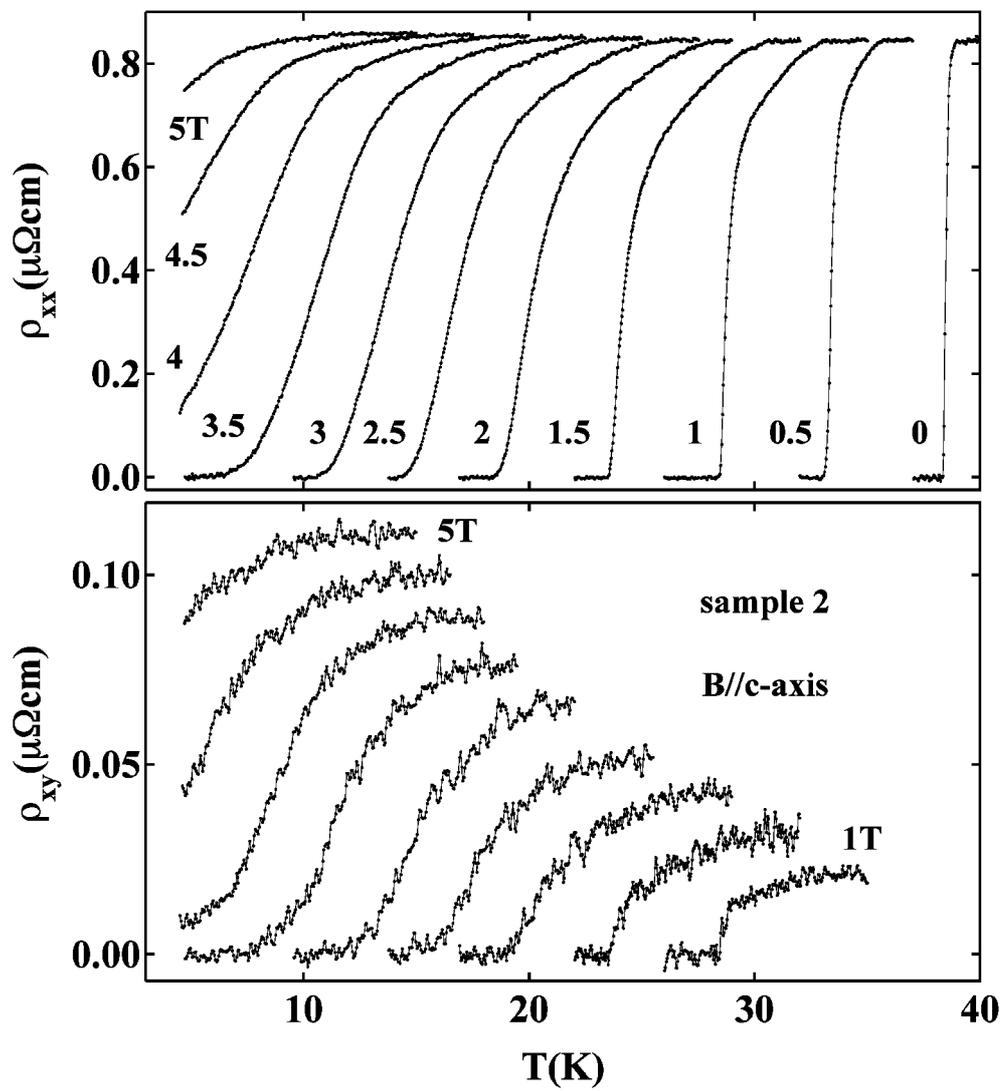

Fig. 3. Temperature dependence of the in-plane longitudinal and Hall resistivity in the mixed state (top panel and bottom panel, respectively) at various magnetic fields up to H=5T as indicated in the Figure.